\documentclass[12pt]{iopart}
\usepackage[active]{srcltx}
\usepackage{graphicx}
\usepackage{iopams}
\usepackage{bm}
\usepackage{mathtext}

\begin{document}

\title{Phase diagrams of the Bose-Fermi-Hubbard model at finite temperature}

 \author{T S  Mysakovych}

  \address{Institute for Condensed Matter Physics,1 Svientsitskii Str., 29011 Lviv, Ukraine}

\ead{mysakovych@icmp.lviv.ua}

\begin{abstract}

The phase transitions at finite temperatures in the systems described by the Bose-Fermi-Hubbard model are investigated in this work in the framework of the selfconsistent random phase approximation. The case of the hard-core bosons is considered and the pseudospin formalism is used. 
 The density-density correlator is calculated in the random phase approximation and the possibilities of transitions from superfluid to supersolid phases are investigated. It is shown that the transitions between  uniform and charge ordered phases can be of the second or the first order, depending on the system parameters.

\end{abstract}

\submitto{\JPCM}

\maketitle

\section{Introduction}

Properties of the systems of ultracold atomic gases confined in optical lattices   are intensively studied in the last years both theoretically and experimentally \cite{jaksch,greiner,gunter,ospelkaus,albus}.  Special attention is paid to the mixture of bosons and spin polarised fermions (e.g., $^6$Li- $^7$Li, $^{40}$K- $^{87}$Rb, $^6$Li- $^{87}$Rb atoms).  Such systems can be well described by the Bose-Fermi-Hubbard model (BFHM) \cite{albus}, which is an extension of the Bose-Hubbard model. 
 The BFHM can also be considered as a generalisation of the fermionic Hubbard model.
It is known for the case of the  Bose-Hubbard model that competition between two terms, one is connected with the on-site energy $U$ and another describes the nearest-neighbour
 hopping with the tunnelling parameter $t$, 
defines 
 the state of the
system (when the kinetics energy dominates the ground state of the system is superfluid, in the opposite case the ground state is a Mott insulator) \cite{fisher,krishnamurthy}. For the case of the BFHM  phase diagrams are more complicated because due to the presence of fermions the effective interaction between bosons is generated.

  The Bose-Fermi mixtures in optical lattices have been studied using a variety of methods  \cite{blatter,lewenstein,cramer,varney,lutchyn,refael,freericks,batrouni,titvinidze,mering2}. In  \cite{blatter}, it was demonstrated that a two-dimensional mixture of bosons and fermions develops a supersolid phase (this phase is characterized by the simultaneous presence of a density wave and phase order in condensate). The case of small fermion hopping was investigated in  \cite{lewenstein} in the framework of composite fermion approach and composite fermions were formed by a fermion and one or several bosons (bosonic holes) for attractive   (repulsive) Bose-Fermi-interactions.  
In  \cite{cramer,varney}, inhomogeneous (due the the presence of the trapping potential) mixtures of bosons and fermions were studied. Enhancement of the superfluidity due to the presence of  fermions  was predicted in  \cite{lutchyn}. The existence of the supersolid phase was confirmed in  \cite{batrouni} using quantum Monte-Carlo simulations.  
A mixture of  the mean field approximation for a bosonic part and the dynamical mean field theory for a fermionic part of the Hamiltonian
 was applied  in  \cite{titvinidze} and  the presence of a supersolid phase  at weak Bose-Fermi interaction   
 was established. The case of 1D Bose-Fermi-Hubbard model (BFHM) in the limit of large fermion hopping was investigated in  \cite{mering2} (the case of half filling was considered only and they did not observe the supersolid state).

It should be noted that the Bose-Fermi-Hubbard-type model can  also be applied for the description of intercalation of ions in crystals
 (for example, lithium intercalation in TiO$_2$ crystals).
Theoretical investigation of such process in most cases were restricted to the numerical ab-initio and density-functional calculations  \cite{stashans,koudriach_2001,koudriach_2002}. 
It was
shown that Li is almost fully ionized once intercalated
 and reconstruction of
electron spectrum at intercalation takes place. Thus, ion-electron 
interaction can play a significant role in such systems.
Another
interesting feature of such crystals is a displacement
of the chemical potential at intercalation into the
conduction band. As a result, these crystals have
metallic conductivity.
At intercalation of lithium in TiO$_2$,
phase separation into Li-poor  and Li-
rich  phases occurrs and  this two-phase
behaviour leads to a constant value of the electrochemical
potential     \cite{wagem_2001,wagem_2003} (this fact is used when constructing
batteries).

In our previous works \cite{my1, my2} we have formulated the pseudospin-electron model of intercalation.
 We have revealed that the effective interaction between bose-atoms (ions) can change 
  its character  
depending on fermionic band filling, that leads to the charge-ordered phase or phase separation into the uniform phases with different concentrations of bosons and fermions.
The ion-electron interaction was also considered in \cite{dublenych} at the investigation 
of thermodynamics of the spin-$1$ model
of intercalation (the model was similar to the known  Blume-Emery-Griffiths model), but the electron as well
as ion transfer was not taken into account. Models of pseudospin-electron model type
are widely used in physics of the strongly-correlated electron systems. Application of such
models to high-temperature superconductors allows one to describe thermodynamics of anharmonic oxygen ion
subsystem and explain the appearance of inhomogeneous states and the bistability phenomena (see \cite{stasyuk}).
Models of a lattice gas are also used
at the description of ionic conductors and at the calculation of their conductivity starting from
works of Mahan \cite{mahan} and others \cite{tomoyose,dulepa}.

In this work we consider the hard-core limit (infinite on-site boson-boson interaction) of the BFHM at finite temperature (most previous investigations considered   the zero-temperature case). 
 Our paper is organized as follows. In section \ref{2sect} we present the description of the model and 
give a self-consistent scheme for calculation of the density-density correlator (susceptibility) in the random phase approximation (RPA).
 In section \ref{3sect} we present phase diagrams for different values of the model parameters.
Special attention is paid to the influence of the temperature change on the phase transitions.
 We present our conclusions in section \ref{4sect}.

\section{MODEL AND METHOD  }\label{2sect}

 We consider the BFHM in the hard core limit.  Using  the pseudospin formalism, the Hamiltonian of the model  is written in the following form
 \begin{eqnarray}\label{ham}
\fl H=-\sum_{ij} \Omega_{ij} S^+_i S^-_j 
 - \sum_{ij} t_{ij} c^+_{i} c_{j} 
 + \sum_{i} g S^z_i n_{i}\nonumber\\
- \sum_i\mu n_{i}
-\sum_i hS^z_i.
 \end{eqnarray} 
 The pseudospin variable $S_i^z$ takes two values ($S_i^z=1/2$ when boson 
  is present in a site $i$  and $S_i^z=-1/2$ in the opposite case),
while $c^+_{i}$ and $c_{i}$ are fermionic  creation and annihilation operators,  
      respectively. 
The first and the second terms in equation (\ref{ham}) are responsible for the 
nearest neighbour boson and fermion hopping, respectively;
$g$-term accounts for the boson-fermion interaction energy.
 To control the  number of bosons and fermions we introduce the bosonic and fermionic chemical potentials $h$ and $\mu$,
 respectively.

 The unperturbed Hamiltonian $H_0$ in the mean feald approximation (MFA) is obtained using the following 
 simplification:
 \begin{eqnarray}
   g n_i S^z_i \rightarrow g \langle n_i \rangle S^z_i +
  g n_i \langle  S^z_i \rangle   -
  g \langle  n_i \rangle   \langle  S^z_i \rangle \\
   \Omega S^+_iS^-_j \rightarrow \Omega \langle S^+_i \rangle S^-_j +
  \Omega S^+_i \langle  S^-_j \rangle   -
  \Omega \langle  S^+_i \rangle   \langle  S^-_j \rangle.
\end{eqnarray}
     The Hamiltonain  becomes
 \begin{eqnarray}
   H=H_{ 0}+H_{\rm int},\\
\fl H_{\rm 0}=-\sum_{ ij}t_{ij}c^+_{i} c_{j}+\sum_i\left(g\langle S^z \rangle n_i + gS^z_i \langle n\rangle 
-g \langle S^z \rangle \langle n \rangle\right)\nonumber
\\ -  \sum_i (h S^z_i+ \mu n_i)
 - \sum_{ij} \left(2\Omega_{ij} S^x_i \langle S^x \rangle-\Omega_{ij} \langle S^x \rangle^2\right),\\
\fl H_{\rm int}=\sum_i g\left(S^z_i-\langle S^z \rangle \right)\left(n_i-\langle n \rangle\right)\nonumber
\\- \sum_{ij}\Omega_{ij}\left[(S^x_i-\langle S^x \rangle)(S^x_j-\langle S^x \rangle)
+S^y_iS^y_j\right].
\end{eqnarray}
 It is worth noting that application of the MFA to the strongly correlated systems in the limit of a weak on-site correlation (when there is no correlational splitting of the fermionic band)   allows one to satisfactorily describe their properties.

To diagonalize the Hamiltonian $H_0$ we pass to ${\bi k}$-representation and 
perform the  unitary transformation in the pseudospin subspace:
  \begin{eqnarray}
   S^z_{i}=\sigma^z_{i}\cos\theta+\sigma^x_{i}
   \sin\theta,\\
    S^x_{i   }=\sigma^x_{i  }\cos\theta  -\sigma^z_{i  }
   \sin\theta,\\
    \sin \theta = -\frac{2\Omega \langle S^x \rangle}{{\lambda}  }, 
    \qquad
    \cos\theta  =\frac{h-gn  }{{\lambda}  },\\
    {\lambda}_{  }=\sqrt{(g\langle n\rangle-h)^2+(2 \Omega \langle S^x \rangle )^2}, 
 \qquad   \Omega\equiv \Omega_{{\bi q }=0},\\
\fl H_{\rm 0}=-\sum_{\bi k} (t_{\bi k}+\mu) c^{+}_{\bi k} c_{\bi k}- \sum_i \lambda \sigma_i\nonumber\\-
N g\langle S^z \rangle 
 \langle n \rangle+N\Omega \langle S^x \rangle^2,  
      \end{eqnarray}
where $N$ is the number of lattice sites.

To calculate the density-density correlator  $\mathfrak{G}_{ij}(\tau)=\langle T_\tau S^z_i(\tau) S^z_j(0) \rangle$,   
 we perform an expansion in powers of $H_{\rm int}$ 
\begin{eqnarray}
   \langle T_\tau S^z_i(\tau) S^z_j(0) \rangle=\frac{\langle T_\tau S^z_i(\tau) S^z_j(0) \sigma(\beta)\rangle_0}
{\langle \sigma(\beta)  \rangle_0}, \\ 
 \exp{(-\beta H)}=\exp{(-\beta H_0)}\sigma(\beta),  \\
\sigma(\beta)=T_{\tau}\exp\left[-\int^\beta_0 H_{\rm int}(\tau)\rmd\tau\right], \\
\fl \langle T_\tau S^z_i(\tau) S^z_j(0)  \rangle=
\langle T_\tau S^z_i(\tau) S^z_j(0)  \rangle_0-\nonumber\\
\frac{1}{1!}{\int_0}^{\beta}\rmd\tau_1\langle T_\tau S^z_i(\tau) S^z_j(0) H_{\rm int}(\tau_1) \rangle_0+ ...,
      \end{eqnarray}
the averaging $\langle ... \rangle_0$ is performed over the distribution with $H_0$, 
where $T_{\tau}$ is the imaginary time ordering operator and $\beta=1/T$ is the inverse temperature.

To calculate the average values of the $T_{\tau}$-products of the pseudospin and fermion operators,
 we utilize  the diagrammatic technique 
 based on  Wick's theorem for the spin operators
\cite{izyumov} (besides the usual procedure for the Fermi operators). 
 After elimination in this way of the nondiagonal $\sigma^\pm$ operators we perform 
 the semi-invariant expansion in order 
to calculate the mean values of the
 remaining products 
  of the $\sigma^z$ operators. 
At the summation of diagrams we restrict ourselves to the diagrams having a structure of multi-loop chains in the spirit
 of  the  random phase approximation  (see \cite{my_cmp2002}).  
 The junctions between bosonic (pseudospin) Green's functions and semi-invariants are realised by 
bosonic hopping $\Omega_{\bi q}$
 and the fermionic loop $\Pi_{\bi q}(\omega)$. 
 It is useful to introduce  unperturbed bosonic and fermionic Green's functions
%\begin{eqnarray}
 $ \langle T_\tau \sigma^+_l (\tau) \sigma^-_m(0) \rangle_0=
 -2 \langle \sigma^z \rangle K_{lm} (\tau)$ and  
 $\langle T_\tau c_{\bi k} (\tau)
  c^+_{\bi q}(0) \rangle_0$, respectively, 
%= - \delta_{\bi k, \bi q}G^0_{\bi k} (\tau)$
%\end{eqnarray}
and semi-invariant
%\begin{eqnarray}
  $\langle T_\tau \sigma^z_l (\tau) \sigma^z_m(0) \rangle_0=
 \langle \sigma^z \rangle^2 + M_{lm}$.
%\end{eqnarray}

\begin{figure}[!h]
\centerline{\includegraphics[width=6cm]{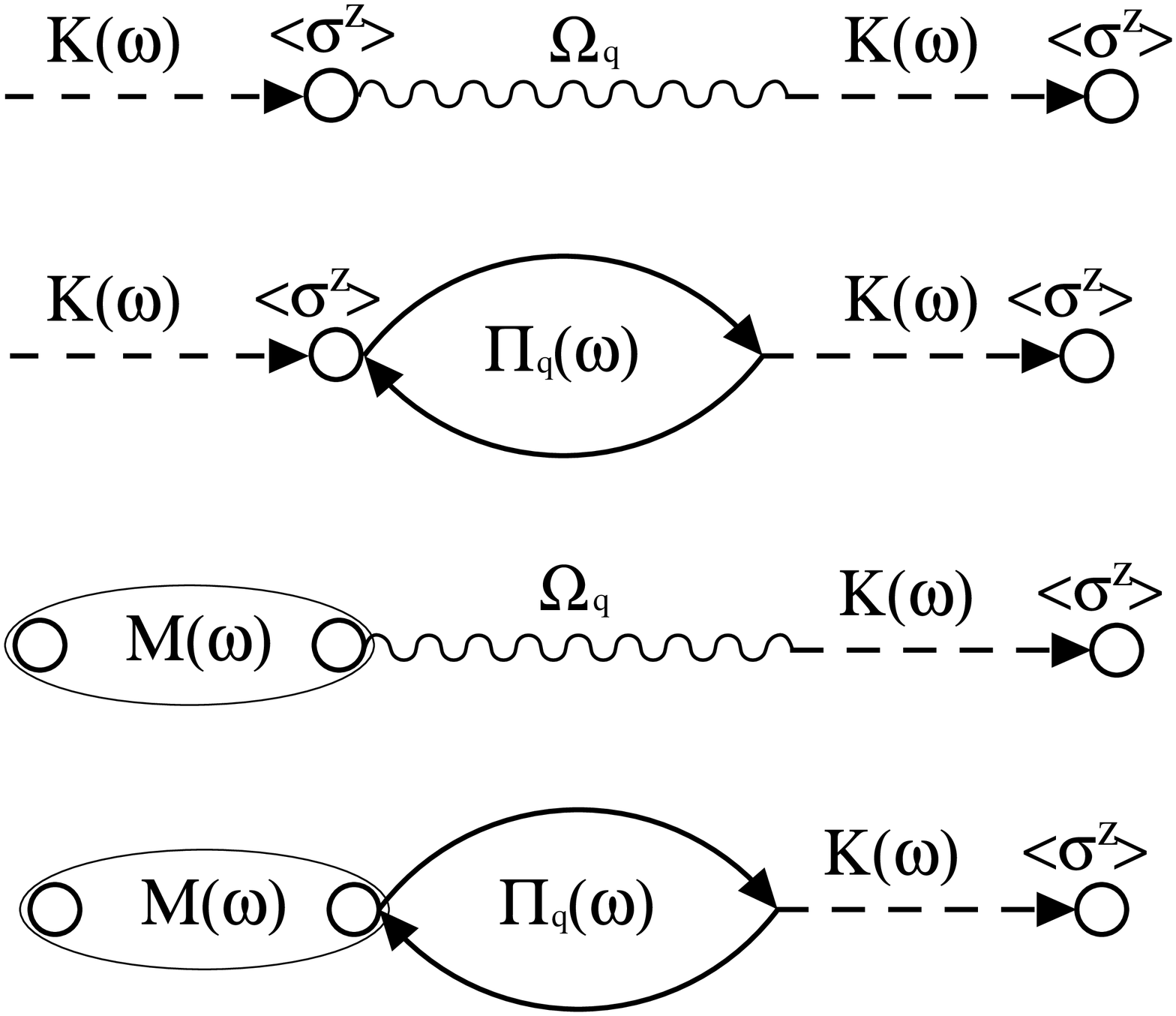} }
\caption{ 
 Typical RPA diagrams  for the   
Green's function $G^{\alpha\beta}_{\bi q}(\omega)$.
 Solid and dashed lines with arrows denote the unperturbed bosonic and fermionic Green's functions, respectively. Wavy lines indicate the  energy dispersion for the bosons $\Omega_{\bi q}$, circles and ovals denote the average value  $\langle \sigma^z \rangle$ and semi-invariants, respectively.
}
\label{01}
\end{figure}

 Let us consider the Green's function  $G^{\alpha\beta}_{lm}(\tau)=
-\frac{1}{2}
\langle T_{\tau} \sigma^{\alpha}_l(\tau) \sigma^{\beta}_{m}(0)  \rangle $  
(with $\alpha,\beta = +,-,z$).
Typical RPA diargams  
 for this Green's function 
$G^{\alpha\beta}_{\bi q}(\omega)$
 in the  frequency representation 
%$G^{\alpha\beta}_{\bi q}(\omega)=
%\int^{\beta}_0 \rmd \tau \rme^{i\omega_n} G^{\alpha\beta}_{\bi q}(\tau) $  (with $\alpha,\beta = +,-,z$),   
are shown in figure~\ref{01}. 
%\begin{eqnarray}
%&&\includegraphics[width=5cm]{dia1.eps}\nonumber\\
%&&\includegraphics[width=5cm]{dia2.eps}\nonumber\\
%&&\includegraphics[width=5cm]{dia3.eps}\nonumber\\
%&&\includegraphics[width=5cm]{dia4.eps},
%\end{eqnarray}
We used the notations for the unperturbed bosonic Green's function  
\begin{eqnarray}
K(\omega_n)=\frac{1}{\rmi\omega_n-\lambda},
\end{eqnarray}
 the fermionic loop 
\begin{eqnarray}
\Pi_{\bi q}(\omega_n)=\frac{1}{N}\sum_{\bi k}\frac{n(t_{\bi k -\bi q})-n(t_{\bi k})}{\rmi\omega_n+t_{\bi k}-t_{\bi k- \bi q}},
\end{eqnarray}
  semi-invariant $M(\omega_n)=\beta\delta_{\omega_n,0}(\frac{1}{4}-\langle\sigma^z\rangle^2)$ and
average value of the pseudospin variable $\langle \sigma^z \rangle=\frac{1}{2}\tanh({\beta\lambda}/{2})$. 

The equation for this Green's function $G^{\alpha\beta}_{\bi q}(\omega_n)$ 
 is
\begin{eqnarray}\label{larkin}
   G^{\alpha\beta}_{\bi q}(\omega_n)=
G^{\alpha\beta}_{(0)\bi q}(\omega_n)\Delta^{\alpha\beta}+G^{\alpha\delta}_{(0)\bi q}(\omega_n)
\Sigma^{\delta\gamma}_{\bi q}(\omega_n) G^{\gamma\beta}_{\bi q}(\omega_n),
      \end{eqnarray}
 where $\Sigma^{\alpha\beta}_{\bi q}(\omega_n)=
\Pi^{\alpha\beta}_{\bi q}(\omega_n)+\Omega^{\alpha\beta}_{\bi q}$ (with $\alpha,\beta = +,-,z$) and $\omega_n$ is the bosonic Matsubara frequency.
These matrix equations (\ref{larkin}) form three independent sets  of equations of the third order and can be separately solved.
 For the case of the Green's functions 
 $G^{+-}_{\bi q}(\omega_n)$, $G^{--}_{\bi q}(\omega_n)$, $G^{z-}_{\bi q}(\omega_n)$ the matrices $\Pi^{\alpha\beta}_{\bi q}(\omega_n)$,  $\Omega^{\alpha\beta}_{\bi q}$  and the unperturbed Green's functions 
 $G^{\alpha\beta}_{(0) \bi q}(\omega_n)$
are 
\begin{eqnarray}
\Pi^{-+}_{\bi q}(\omega_n)=\Pi^{+-}_{\bi q}(\omega_n)=\Pi^{++}_{\bi q}(\omega_n)\\
=\Pi^{--}_{\bi q}(\omega_n)=g^2\Pi_{\bi q}(\omega_n)\frac{\sin^2\theta}{2}, \\
%\end{eqnarray}
%\begin{eqnarray}
 \Pi^{-z}_{\bi q}(\omega_n)=\Pi^{+z}_{\bi q}(\omega_n)= g^2\Pi_{\bi q}(\omega_n)\sin\theta \cos\theta,\\
 \Pi^{z-}_{\bi q}(\omega_n)=\Pi^{z+}_{\bi q}(\omega_n)= -g^2\Pi_{\bi q}(\omega_n)
\frac{\sin\theta \cos\theta}{2},\\
\Pi^{zz}_{\bi q}(\omega_n)= -g^2\Pi_{\bi q}(\omega_n) \cos^2\theta,
  \qquad  \Omega^{zz}_{\bi q}= 2\Omega_{\bi q} \sin^2\theta,
 \\
\Omega^{-+}_{\bi q}=\Omega^{+-}_{\bi q}=
-\Omega_{\bi q}(1+\cos^2\theta), \\ \Omega^{++}_{\bi q}
=\Omega^{--}_{\bi q}=-\Omega_{\bi q}(\cos^2\theta-1),\\
 \Omega^{-z}_{\bi q}=\Omega^{+z}_{\bi q}=2\Omega_{\bi q}\sin\theta \cos\theta,
  \\ \Omega^{z-}_{\bi q}=\Omega^{z+}_{\bi q}=-\Omega_{\bi q}
\sin\theta \cos\theta, \\
G^{+-}_{(0)}(\omega_n)=K(\omega_n)\langle \sigma^z  \rangle, \qquad 
G^{--}_{(0)}(\omega_n)=K(-\omega_n)\langle \sigma^z  \rangle, \\
G^{z-}_{(0)}(\omega_n)=M(\omega_n), \qquad
\Delta^{+-}=1, \qquad \Delta^{--}=0, \qquad 
\Delta^{z-}=0. 
\end{eqnarray}
Similar matrix equations can be written for the Green's functions 
 $G^{++}_{\bi q}(\omega_n)$, $G^{-+}_{\bi q}(\omega_n)$, $G^{z+}_{\bi q}(\omega_n)$ and 
 $G^{+z}_{\bi q}(\omega_n)$, $G^{-z}_{\bi q}(\omega_n)$, $G^{zz}_{\bi q}(\omega_n)$ with the corresponding matrices
 $\Pi^{\alpha\beta}_{\bi q}(\omega_n)$ and $\Omega^{\alpha\beta}_{\bi q}$ (we do not present here these matrices).
As a result, we can solve these three sets of equations of the third order and after some tedious algebra we derive the expression 
 for the density-density correlator
\begin{eqnarray}\label{susceptibility1}
 \fl \mathfrak G_{ij}(\tau)=\langle T_\tau \sigma^z_i(\tau) \sigma^z_j(0) \rangle \cos^2\theta +
\langle T_\tau \sigma^x_i(\tau) \sigma^x_j(0) \rangle \sin^2\theta\nonumber\\
+
\langle T_\tau \sigma^z_i(\tau) \sigma^x_j(0) \rangle \sin\theta\cos\theta+
\langle T_\tau \sigma^x_i(\tau) \sigma^z_j(0) \rangle \sin\theta\cos\theta,
\end{eqnarray}
\begin{eqnarray}\label{susceptibility}
\fl \mathfrak G_{\bi q}(\omega_n)=
 \frac{ \sin^2\theta \langle \sigma^z \rangle+\lambda M(\omega_n) \cos^2\theta-
         2\Omega_{\bi q}M(\omega_n)\langle\sigma^z  \rangle 
       }{\mathfrak \Delta}\nonumber\\
\times(\lambda-2\langle \sigma^z \rangle \Omega_{\bi q}),
\\
\fl 
{\mathfrak \Delta}= -(\rmi\omega_n)^2+(\lambda-2 \langle\sigma^z \rangle \Omega_{\bi q} )
 \big[  \lambda-2\langle \sigma^z \rangle\cos^2\theta\Omega_{\bi q}\nonumber\\ 
+ \langle\sigma^z \rangle g^2 \sin^2\theta
\Pi_{\bi q}(\omega_n)
 -2M(\omega_n)\Omega_{\bi q}\lambda \sin^2\theta\nonumber\\
+M(\omega_n)g^2\lambda\Pi_{\bi q}(\omega_n)\cos^2\theta-
2\langle \sigma^z \rangle\Omega_{\bi q}M(\omega_n)\Pi_{\bi q}(\omega_n)g^2 \big].
      \end{eqnarray}

If we use the equation of motion method developed for the two-time Green's function $\langle\langle S^z(t) | S^z(t')  \rangle\rangle=-\rmi\theta(t-t')\langle[S^z(t),S^z(t')]\rangle$ and decoupling in the spirit of Tyablikov approximation
%\begin{eqnarray}\label{tyablik}
 $[\sigma^x_i,H]\approx -2 \rmi \langle \sigma^z \rangle\sum_l\sigma^y_l\Omega_{il}+
\rmi \lambda \sigma^y_i$
%\end{eqnarray}
%we  obtain the equations
%\begin{eqnarray}
%\fl \omega\langle\langle \sigma^x|\sigma^x \rangle\rangle_{\bi q, \omega}=\rmi \lambda  
%\langle\langle \sigma^y|\sigma^x \rangle\rangle_{\bi q, \omega}-
%2 \rmi \langle \sigma^z \rangle 
% \Omega_{\bi q}
%\langle\langle \sigma^y|\sigma^x \rangle\rangle_{\bi q, \omega},\\
%\fl \omega\langle\langle \sigma^y|\sigma^x \rangle\rangle_{\bi q, \omega}=
%-\rmi\langle\sigma^z \rangle-
%\rmi \lambda  
%\langle\langle \sigma^x|\sigma^x \rangle\rangle_{\bi q,\omega}\\+
%2 \rmi \langle \sigma^z \rangle \cos^2\theta
% \Omega_{\bi q}
%\langle\langle \sigma^x|\sigma^x \rangle\rangle_{\bi q,\omega}
%-
%\rmi \langle \sigma^z \rangle\sin\theta g 
%\langle\langle n|\sigma^x \rangle\rangle_{\bi q,\omega}\nonumber,\\
%\fl \langle\langle n|\sigma^x \rangle\rangle_{\bi q,\omega}=
%g \sin \theta \Pi_{\bi q}(\omega)\langle\langle \sigma^x|\sigma^x \rangle\rangle_{\bi q, \omega}.
%\end{eqnarray}
%As a result, we can write 
%\begin{eqnarray}\label{twotime}
%\fl \langle\langle S^z | S^z  \rangle\rangle_{\bi q, \omega}=
%\frac{ \sin^2\theta \langle \sigma^z \rangle (\lambda-2\langle \sigma^z \rangle \Omega_{\bi q})}
% {\mathfrak D 
%},\\
%\fl \mathfrak D
%=\omega^2-(\lambda-2 \langle\sigma^z \rangle \Omega_{\bi q} )\nonumber\\
% \times\left(\lambda-2\langle \sigma^z \rangle\cos^2\theta\Omega_{\bi q} +
% \langle\sigma^z \rangle g^2 \sin^2\theta \Pi_{\bi q}(\omega)\right).
%      \end{eqnarray}
 we can obtain the expression for the correlator $\langle\langle S^z | S^z  \rangle\rangle_{\bi q, \omega}$
 which 
is similar to the equation (\ref{susceptibility})  
but differs from it due to the absence of the  terms proportional to  $\delta_{\omega_n,0}$.  
These terms have appeared in the diagrammatic technique due to the presence of the semi-invariants
 and are important when we investigate the static limit $\omega \rightarrow 0$.
The equation of motion  method does not allow us to take into account these  terms and because of this 
we should  use the diagrammatic technique.
It should be noted that such an peculiarity was also pointed out  in \cite{menchyshyn} at the investigation of the Bose-Hubbard model   in the hard-core case.

\section{PHASE DIAGRAMS}\label{3sect}

Lines of the instability with respect to the transition into the phase with charge ordering can be obtained using the condition of divergence of the static density-density correlator $\mathfrak{G}_{\bi q}(\omega=0)$. We consider two cases: i) the transition from a normal (NR) nonsuperfluid uniform to  nonsuperfluid charge-density-wave (CDW) phase ii) the transition from a superfluid phase to superfluid phase with long-range ordering  (a supersolid phase). 
The equations for averages $\langle n \rangle$, $\langle S^z \rangle$, $\langle S^x \rangle$ 
are obtained in the mean field approximation. 
Let us introduce two sublattices: 
     $\langle   n_{i\alpha}  \rangle=n_\alpha, \ \ 
     \langle S^z_{i\alpha} \rangle= \langle S^z_{\alpha} \rangle $, 
 $\alpha=1,2$ is a sublattice index, $i$ is an elementary cell index. Using the Hamiltonian $H_0$, we can obtain the following equations for averages
\cite{my2}
\begin{eqnarray}
\label{n_mod}
      \fl n_{\alpha}{=}\frac{1}{N}\sum_{\bi{k} }\frac{1+\cos( 2 \phi)}{2}
 \left(\rme^{\frac{\lambda_{\bi{k} \alpha}-\mu}{T}}+1\right)^{-1}\nonumber
\\+\sum_{\bi{k}}\frac{1-\cos (2 \phi)}{2}
 \left(\rme^{\frac{\lambda_{\bi{k} \beta}-\mu}{T}}+1\right)^{-1},\\  \label{n_mod2}
        \langle S^z_{\alpha} \rangle=\frac{h-gn_{\alpha}}{2
       \tilde{\lambda}_{\alpha}}
 \tanh\left(\frac{\beta \tilde{\lambda}_{\alpha}}{2}\right), \\ \label{n_mod3}
 \langle S^x_\alpha \rangle=\frac{2\Omega \langle S^x_\beta \rangle  }
{2\tilde{\lambda}_\alpha}\tanh\left(\frac{\beta \tilde{\lambda}_{\alpha}}{2}\right)
\end{eqnarray}
with
\begin{eqnarray}
\lambda_{\bf{k}\alpha}=g\frac{\langle S^z_{1} \rangle+\langle S^z_{2} \rangle}
{2}+(-1)^{\alpha} \sqrt{\left(g\frac{ \langle S^z_{1} \rangle  -  \langle S^z_{2} \rangle }{2}\right)^2+t^{2}_{\bf{k}}},
 \\ \sin2\phi=\frac{-t_{\bf{k}}}{\sqrt{\left(g\frac{\langle S^z_{1} \rangle
    -\langle S^z_{2} \rangle     }{2}\right)^2+
t^{2}_{\bf{k}}}},\\  
\tilde{\lambda}_{\alpha}=\sqrt{(gn_{\alpha}-h)^2+\left(2 \Omega \langle S^x_\beta \rangle \right)^2},\ \ \
\alpha\neq\beta.
      \end{eqnarray}
The grand canonical potential can be written as \cite{my2}
\begin{eqnarray}
\label{Potent_mod}
  \fl \frac{ \Phi}{\frac{N}{2}}=-\frac{T}{N}\sum_{\bi{k}}
      \ln\left[\left(1+\rme^{\frac{\mu-\lambda_{\bi{k}1}}{T}}\right)
      \left(1+\rme^{\frac{\mu-\lambda_{\bi{k}2}}{T}}\right)\right]\nonumber\\
      -T\ln\left[4\cosh\left(\frac{\beta\tilde{\lambda}_1}{2}\right)
      \cosh\left(\frac{\beta\tilde{\lambda}_2}{2}\right)\right]\nonumber\\
-g\left(n_1 \langle S^z_{1} \rangle +n_2  \langle S^z_{2} \rangle \right)
 +2\Omega \langle S^x_1 \rangle \langle S^x_2 \rangle.
              \end{eqnarray}
The doubling of the unit cell leads to the splitting in the fermionic spectrum with the gap 
 $g|\langle S^z_{1} \rangle-\langle S^z_{2} \rangle|$. 
The differences $\delta n=\langle n_1 \rangle - \langle n_2 \rangle$, 
$\delta S^z=\langle S^z_1 \rangle - \langle S^z_2 \rangle$, 
$\delta S^x=\langle S^x_1 \rangle - \langle S^x_2 \rangle$
play the role of the order parameter for the modulated phase ($\langle S^x \rangle\neq 0$  in the superfluid phase and   $\delta S^x\neq 0$ in the supersolid phase). Coming from the set of equations (\ref{n_mod}), (\ref{n_mod2}) and (\ref{n_mod3}), we can write the equations for $\delta n$, $\delta S^z$, $\delta S^x$ and separate the contributions of the first order. As a result, we obtain the condition of the appearance of nonzero solutions for $\delta n$, $\delta S^z$ and $\delta S^x$. It can be shown that this condition coincides with the condition when the static  density-density correlator  
$\mathfrak{G}_{ q=\pi}(\omega=0)$ diverges. Therefore our scheme for calculation of the density-density correlator in the RPA and the corresponding averages  $\langle n \rangle$, 
$\langle S^z\rangle$  and $\langle S^x \rangle$ in the MFA is  a self-consistent scheme.

\begin{figure}[!h]
\centerline{\includegraphics[width=6.5cm]{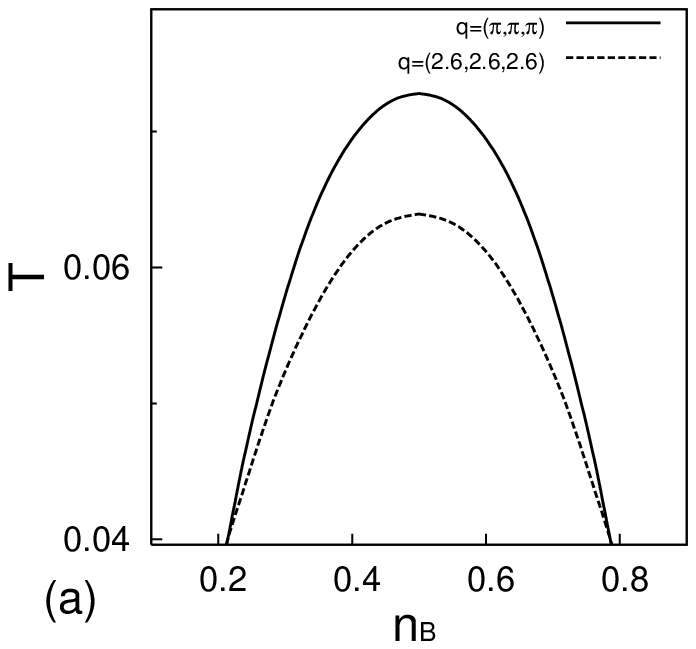}  
\includegraphics[width=6.5cm]{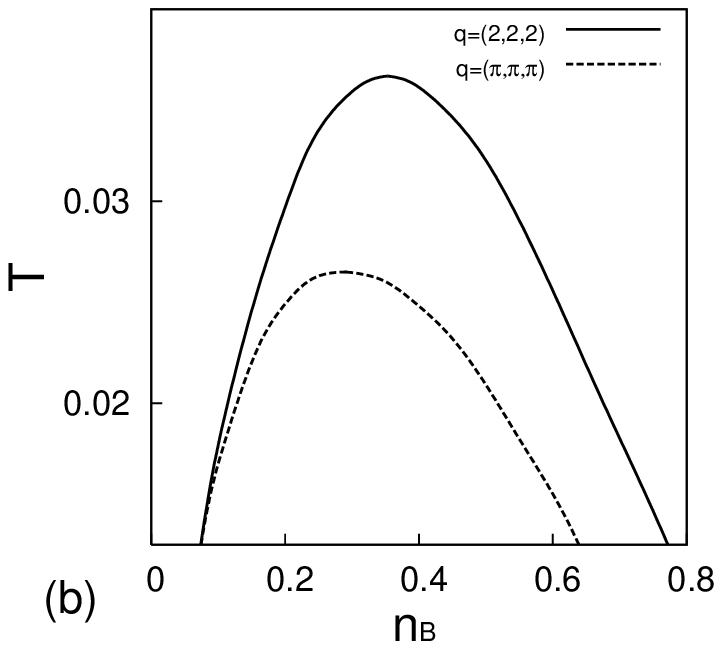}}
\caption{ 
The lines of  
 instability of the nonsuperfluid phase with respect to the transition into the charge ordered phase 
  for    
$W=1$, $g=-0.4$,    $\Omega=0$, $\mu=0$ (a) and $\mu=0.3$ (b).}
\label{1}
\end{figure}

At the numerical calculations of the density-density correlator we  consider a 
 three-dimensional case with a lattice constant $a=1$ and in our calculations we choose a half width of the fermionic band $W$ to be our  energy scale 
($-W<t_{\bi k}<W$). First we investigate the uniform phase ($\langle n_1 \rangle=\langle n_2 \rangle$). From the set of equations (\ref{n_mod}), (\ref{n_mod2}) and (\ref{n_mod3})  it follows  that the solutions of these equations with $\langle S^x \rangle\neq 0$ can be realised when $\Omega>2T$. Therefore at finite temperature we can consider the transition from the normal uniform nonsuperfluid  phase (at low temperatures this is a Mott insulating  phase) to the CDW phase for small values of the bosonic hopping parameter ($\Omega<2T$).
 In figure~\ref{1}(a), we plot  lines of the instability with respect to the transition into the charge ordered phase at the fixed
 fermionic chemical potential  (the case $\mu=0$ corresponds to the  half-filling case) for the
case of the nonsuperfluid phase
 (the bosonic concentration $n_B=S^z+1/2$). As seen in figure~\ref{1}(a), the highest temperature of the instability is realised for the case of the chess-board phase with the wave vector 
$\bi q=(\pi,\pi,\pi)$. In figure~\ref{1}(b)   the lines of the instability for the case $\mu=0.3$ (when the system goes away from the half-filling case) are plotted.
 From figure~\ref{1}(b) we observe that the incommensurate charge ordered phase with the wave vector ${\bi q\approx(2,2,2)}$ has the highest temperature of the instability
 and the system undergoes the transition to the incommensurate modulated phase.
It should be noted that  the condition of the divergence of the static density-density correlator allows one to investigate the phase transitions of the second order only.

\begin{figure}[!h]
\centerline{\includegraphics[width=6.5cm]{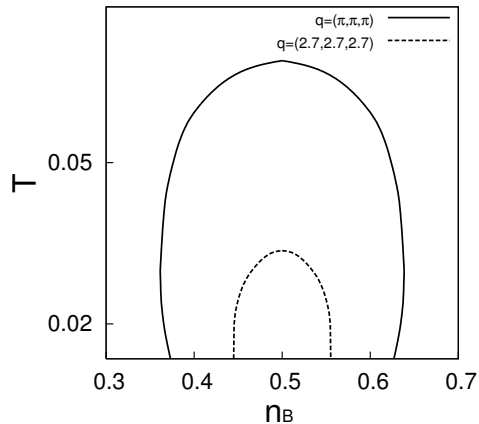}  
}
\caption{ 
  The lines of   instability of the superfluid phase 
 with respect to the transition into the supersolid phase
  for  
$W=1$, $g=-0.4$,    $\Omega=0.15$ and $\mu=0$.}
\label{2}
\end{figure}

Now let us consider the transition to the supersolid phase. In figure~\ref{2}  
 lines of instability of the superfluid phase with respect to the transition into the supersolid phase for the half-filling case $\mu=0$, $\Omega=0.15$ are depicted. 
 As shown in figure~\ref{2},  the transition  to the supersolid phase with modulation wave vector 
$\bi q=(\pi,\pi,\pi)$ is realised. 
We revealed that when the system goes away from the half-filling case and $\mu\neq 0$ 
 the supersolid phase with the modulation wave vector $\bi q=(\pi,\pi,\pi)$ also has  the highest temperature of the transition.

It should be  emphasised that appearance of the CDW and supersolid phases is connected with the presence of the effective interaction between bosons which is formed due to the boson-fermion correlation.
 This interaction  depends on the filling of the fermionic band.

\begin{figure}[htb!]
\centerline{\includegraphics[width=6.5cm]{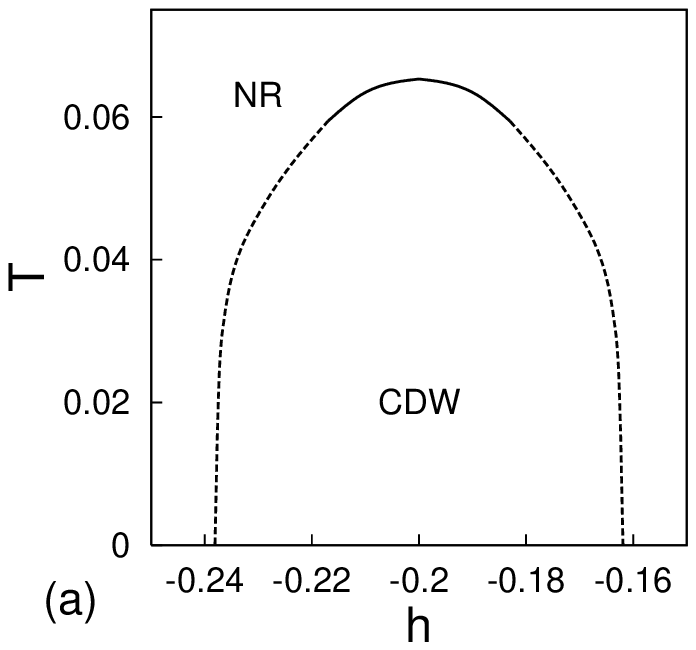} \ \ 
\includegraphics[width=6.5cm]{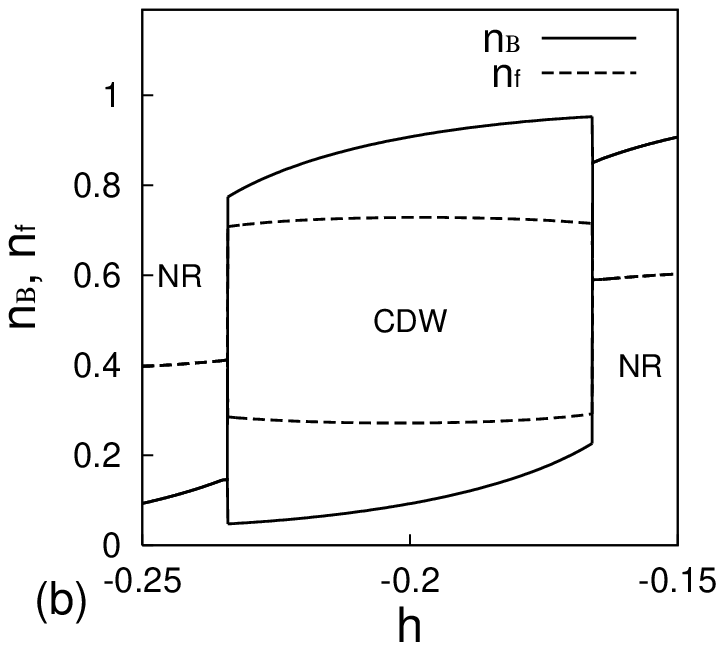}}
\caption{ 
 (a) The lines of the phase transitions of the second  (solid lines) and  first  (dashed lines) order for 
$W=1$, $g=-0.4$,     $\mu=0$ and $\Omega=0$. (b) The dependence of the bosonic and fermionic concentrations on the bosonic chemical potential at $T=0.04$. }
\label{3}
\end{figure}

\begin{figure}[htb!]
\centerline{\includegraphics[width=6.5cm]{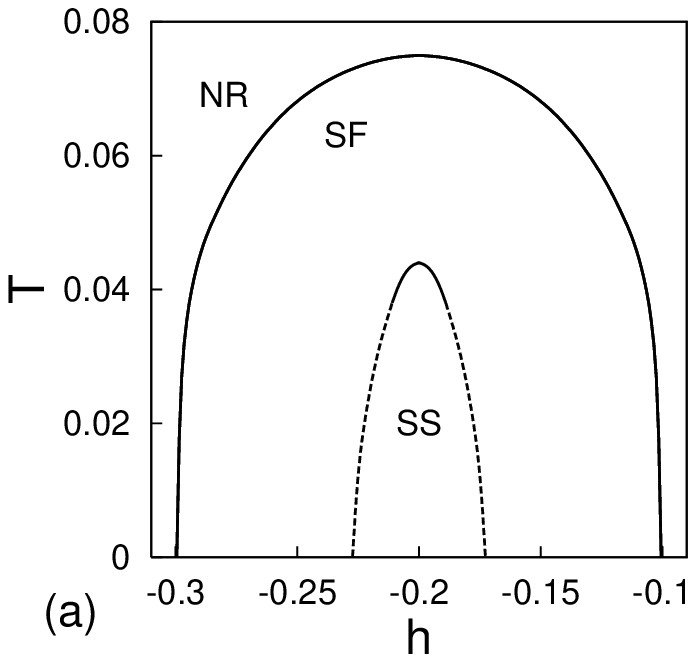} \ \ 
\includegraphics[width=6.5cm]{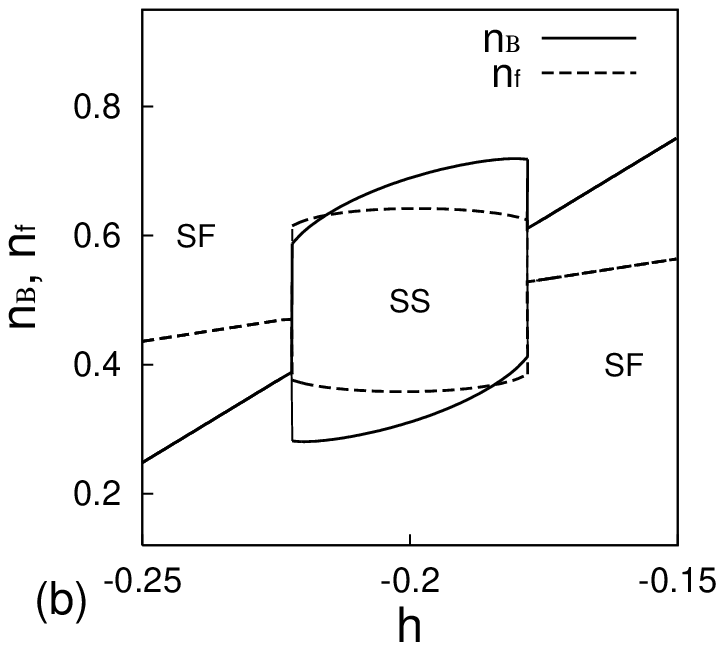}}
\caption{ 
 (a) The lines of the phase transitions of the second  (solid lines) and  first  (dashed lines) order for 
$W=1$, $g=-0.4$,     $\mu=0$ and $\Omega=0.15$. 
(b) The dependence of
 the bosonic and fermionic concentrations on the bosonic chemical potential at $T=0.02$. }
\label{3r}
\end{figure}

Now we want to investigate
the case of 
 the chess-board phase in more detail. 
We use the equations for averages (\ref{n_mod}), (\ref{n_mod2}), (\ref{n_mod3}) and the expression for the grand canonical potential (\ref{Potent_mod}) to find thermodynamically stable states (in this part of our numerical calculations we use 
 the semielliptic density of states, $\rho(t)=\frac{2}{\pi W^2}\sqrt{W^2-t^2}$).  The phase transition 
lines and particle concentrations as  functions of the bosonic chemical potential are shown in figure~\ref{3} and figure~\ref{3r}. The phase transition from the normal uniform nonsuperfluid to chess-board phase can be of the second  or  first order, see figure~\ref{3}a. 
The existence of the phase transition of the first order leads to phase separation in the regime of the fixed concentrations into the NR and CDW phases. 
As shown in figure~\ref{3r}, similar picture is obtained for the transition from the superfluid to the supersolid phase and the transition from the superfluid to the supersolid phase can be of the first or second order depending on the system parameters.

\begin{figure}[htb!]
\centerline{\includegraphics[width=6.5cm]{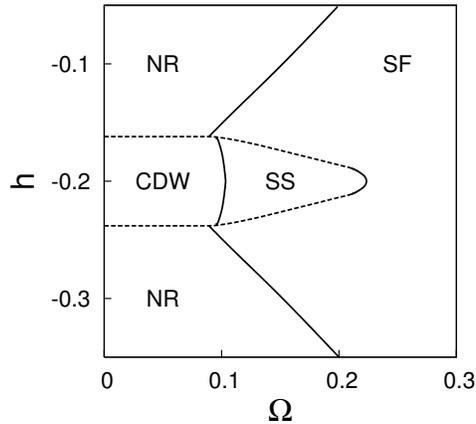}}
\caption{ Phase diagram in the ($h-\Omega$) plane for  
$W=1$, $g=-0.4$,     $\mu=0$ and $T=0.01$. Solid (dashed) lines denote the phase transitions of the second (first) order.}
\label{4}
\end{figure}

In figure~\ref{4}, we show the phase diagrams in the plane ($h-\Omega$) at low temperature. 
As temperature increases, the regions of the existence of the CDW phase and the supersolid phase
 are possible for smaller parameter space and the first order phase transitions from the normal uniform nonsuperfluid (superfluid) into the CDW (SS) phases transforms into the second one.
 It should be noted that similar diagrams at $T=0$ were obtained in \cite{mering2}, but they did not reveal the possibility of the transition to the supersolid phase.
% (may be this is connected with the fact that they considered the regime of the fixed bosonic concentration  with %$n_b=\frac{1}{2}$).

 \section{CONCLUSIONS}\label{4sect}

The phase transitions in the Bose-Fermi-Hubbard model at finite temperature has been considered in this work. 
 We  studied the hard-core limit and used pseudospin formalism. The thermodynamics of the model was investigated in the case of the weak boson-fermion interaction. The analytical expression for the density-density correlator 
has been obtained  in the 
 framework of the 
self-consistent scheme of the random phase approximation. The effective boson-boson interaction is formed due to the boson interaction with fermions, this effective interaction depends on the filling of the fermionic band. 
It is revealed that at small values of the bosonic tunnelling amplitude the system undergoes the phase transition from the uniform nonsuperfluid phase to the chess-board phase (the case of  half filling of the fermion band) or to the incommensurate phase (when the system goes away from the half filling) at the lowering of the temperature. At increase of the  bosonic hopping parameter the phase transition from the superfluid phase to the supersolid phase with a doubly modulated lattice period takes place (it should be noted that the presence of the supersolid phase at the weak boson-fermion interaction and zero-temperature was also established in \cite{titvinidze} in the framework of a generalized dynamical mean field theory).  
 The transition from the uniform  to modulated phase can be of the first or  second order,
 depending on the model parameters  and temperature. The presence of the first order phase transition means that in the regime of the fixed fermionic concentrations the phase separation into the   uniform and modulated phases  is possible.

\ack

 I would like to thank I.V. Stasyuk for usefull discussions.

\end{document}